


\font\titlefont = cmr10 scaled\magstep 4
\font\sectionfont = cmr10
\font\littlefont = cmr5
\font\eightrm = cmr8

\def\ss{\scriptstyle}
\def\sss{\scriptscriptstyle}

\newcount\tcflag
\tcflag = 0  

\ifnum\tcflag = 0 \magnification = 1200 \fi  

\global\baselineskip = 1.2\baselineskip
\global\parskip = 4pt plus 0.3pt
\global\abovedisplayskip = 18pt plus3pt minus9pt
\global\belowdisplayskip = 18pt plus3pt minus9pt
\global\abovedisplayshortskip = 6pt plus3pt
\global\belowdisplayshortskip = 6pt plus3pt

\def\barsoff{\overfullrule=0pt}


\def\endignore{}
\def\ignore #1\endignore{}

\newcount\dflag
\dflag = 0


\def\monthname{\ifcase\month
\or January \or February \or March \or April \or May \or June%
\or July \or August \or September \or October \or November %
\or December
\fi}

\newcount\dummy
\newcount\minute  
\newcount\hour
\newcount\localtime
\newcount\localday
\localtime = \time
\localday = \day

\def\advanceclock#1#2{ 
\dummy = #1
\multiply\dummy by 60
\advance\dummy by #2
\advance\localtime by \dummy
\ifnum\localtime > 1440 
\advance\localtime by -1440
\advance\localday by 1
\fi}

\def\settime{{\dummy = \localtime%
\divide\dummy by 60%
\hour = \dummy
\minute = \localtime%
\multiply\dummy by 60%
\advance\minute by -\dummy
\ifnum\minute < 10
\xdef\spacer{0} 
\else \xdef\spacer{}
\fi %
\ifnum\hour < 12
\xdef\ampm{a.m.} 
\else
\xdef\ampm{p.m.} 
\advance\hour by -12 %
\fi %
\ifnum\hour = 0 \hour = 12 \fi
\xdef\timestring{\number\hour : \spacer \number\minute%
\thinspace \ampm}}}



\def\endtitle{}
\def\title#1\endtitle{\vskip.5in\titlefont
\global\baselineskip = 2\baselineskip
#1\vskip.4in
\baselineskip = 0.5\baselineskip\rm}

\def\endauthors{}
\def\authors#1\endauthors{#1}

\def\endabstract{}
\def\abstract#1\endabstract{\vskip .3in%
\centerline{\sectionfont\bf Abstract}%
\vskip .1in
\noindent#1}

\def\nopageonenumber{\footline={\ifnum\pageno<2\hfil\else
\hss\tenrm\folio\hss\fi}}  

\newcount\nsection
\newcount\nsubsection

\def\section#1{\global\advance\nsection by 1
\nsubsection=0
\bigskip\noindent\centerline{\sectionfont \bf \number\nsection.\ #1}
\bigskip\rm\nobreak}

\def\subsection#1{\global\advance\nsubsection by 1
\bigskip\noindent\sectionfont \sl \number\nsection.\number\nsubsection)\
#1\bigskip\rm\nobreak}

\def\topic#1{{\medskip\noindent $\bullet$ \it #1:}}
\def\endtopic{\medskip}

\def\appendix#1#2{\bigskip\noindent%
\centerline{\sectionfont \bf Appendix #1.\ #2}
\bigskip\rm\nobreak}


\newcount\nref
\global\nref = 1

\def\therefs{}


\def\ref#1#2{\xdef #1{[\number\nref]}
\ifnum\nref = 1\global\xdef\therefs{\item{[\number\nref]} #2\ }
\else
\global\xdef\oldrefs{\therefs}
\global\xdef\therefs{\oldrefs\vskip.1in\item{[\number\nref]} #2\ }%
\fi%
\global\advance\nref by 1
}

\def\listrefs{\vfill\eject\section{References}\therefs}


\newcount\nfoot
\global\nfoot = 1

\def\foot#1#2{\xdef #1{(\number\nfoot)}
\footnote{${}^{\number\nfoot}$}{\eightrm #2}
\global\advance\nfoot by 1
}


\newcount\nfig
\global\nfig = 1
\def\thefigs{} 

\def\figure#1#2{\xdef #1{(\number\nfig)}
\ifnum\nfig = 1\global\xdef\thefigs{\item{(\number\nfig)} #2\ }
\else
\global\xdef\oldfigs{\thefigs}
\global\xdef\thefigs{\oldfigs\vskip.1in\item{(\number\nfig)} #2\ }%
\fi%
\global\advance\nfig by 1 } 

\def\fig#1{\xdef #1{(\number\nfig)}
\global\advance\nfig by 1 } 


\newcount\cflag
\newcount\nequation
\global\nequation = 1
\def\eqlabel{(1)}

\def\nexteqno{\ifnum\cflag = 0
\global\advance\nequation by 1
\fi
\global\cflag = 0
\xdef\eqlabel{(\number\nequation)}}

\def\lasteqno{\global\advance\nequation by -1
\xdef\eqlabel{(\number\nequation)}}

\def\label#1{\xdef #1{(\number\nequation)}
\ifnum\dflag = 1
{\escapechar = -1
\xdef\draftname{\littlefont\string#1}}
\fi}

\def\clabel#1#2{\xdef\eqlabel{(\number\nequation #2)}
\global\cflag = 1
\xdef #1{\eqlabel}
\ifnum\dflag = 1
{\escapechar = -1
\xdef\draftname{\string#1}}
\fi}

\def\cclabel#1#2{\xdef\eqlabel{#2)}
\global\cflag = 1
\xdef #1{\eqlabel}
\ifnum\dflag = 1
{\escapechar = -1
\xdef\draftname{\string#1}}
\fi}


\def\eeq{}

\def\eqnn #1\eeq{$$ #1 $$}

\def\eq #1\eeq{
\ifnum\dflag = 0
{\xdef\draftname{\ }}
\fi 
$$ #1
\eqno{\eqlabel \rlap{\ \draftname}} $$
\nexteqno}







\def\eqa #1\eeq{
\ifnum\dflag = 0
{\xdef\draftname{\ }}
\fi 
$$ \eqalignno{ #1 } $$
\global\cflag = 0}


\def\etc{{\it etc.\/}}

\def\apriori{{\it a priori\/}}


\def\npb#1#2#3{{\it Nucl.\ Phys.} {\bf B#1} (19#2) #3}
\def\plb#1#2#3{{\it Phys.\ Lett.} {\bf #1B} (19#2) #3}

\def\prd#1#2#3{{\it Phys.\ Rev.} {\bf D#1} (19#2) #3}

\def\prl#1#2#3{{\it Phys.\ Rev.\ Lett.} {\bf #1} (19#2) #3}

\def\zpc#1#2#3{{\it Zeit.\ Phys.} {\bf C#1} (19#2) #3}


\global\nulldelimiterspace = 0pt



\def\frac#1#2{{{#1} \over {#2}}\,}  



\def\Dsl{\hbox{/\kern-.6700em\it D}} 
\def\dsl{\hbox{/\kern-.5300em$\partial$}}
\def\pxpsl{\hbox{/\kern-.5600em$p$}}
\def\ssl{\hbox{/\kern-.5300em$s$}}
\def\epssl{\hbox{/\kern-.5100em$\epsilon$}}
\def\delsl{\hbox{/\kern-.6300em$\nabla$}}
\def\lxpsl{\hbox{/\kern-.4300em$l$}}
\def\elxpsl{\hbox{/\kern-.4500em$\ell$}}
\def\kxpsl{\hbox{/\kern-.5100em$k$}}
\def\qxpsl{\hbox{/\kern-.5000em$q$}}
\def\sla#1{\raise.15ex\hbox{$/$}\kern-.57em #1}
\def\Pl{\gamma_{\sss L}}
\def\Pr{\gamma_{\sss R}}



\def\roughly#1{\mathrel{\raise.3ex\hbox{$#1$\kern-.75em\lower1ex\hbox{$\sim$}}}}





\def\Scl{{\cal L}}


\def\ssl{{\sss L}}

\def\ssq{{\sss Q}}
\def\ssr{{\sss R}}

\def\ssz{{\sss Z}}







\def\GeV{{\rm \ GeV}}


\let\nopictures=Y

\nopageonenumber
\baselineskip = 16pt
\barsoff


\def\pbinv{pb^{-1}}
\def\bk{\item{}}
\def\sw{s_w}
\def\cw{c_w}
\def\ssq{s^2_w}

\def\deg{\delta g}

\def\tauL{\deg^{\tau}_\ssl}

\def\tauR{\deg^{\tau}_\ssr}

\def\UD{\delta_{UD}}


\def\AeFB{A_{\sss FB}(e)}
\def\AmuFB{A_{\sss FB}(\mu )}
\def\AtauFB{A_{\sss FB}(\tau)}
\def\AbFB{A_{\sss FB}(b)}
\def\AcFB{A_{\sss FB}(c)}
\def\Gtot{\Gamma_\ssz}
\def\Sigma{\sigma^h_p}

\def\Rel{R_e}
\def\Rmu{R_\mu}
\def\Rtau{R_\tau}
\def\Rb{R_b}
\def\Rc{R_c}
\def\AePtau{A_e(P_\tau )}
\def\MnsPtau{A_{pol}(\tau )}


\line{hep-ph/9502402 \hfil McGill-95/07, NEIP-95-002, UTTG-06-95}
\rightline{February, 1995.}

\title
\centerline{The Sensitivity to New Physics of a}
\centerline{LEP Scan in 1995}
\endtitle
\authors
\centerline{P. Bamert${}^a$, C.P. Burgess${}^b$ and I. Maksymyk${}^c$}
\vskip .15in
\centerline{\it ${}^a$ Institut de Physique, Universit\'e de Neuch\^atel}
\centerline{\it 1 Rue A.L. Breguet, CH-2000 Neuch\^atel, Switzerland.}
\vskip .1in
\centerline{\it ${}^b$ Physics Department, McGill University}
\centerline{\it 3600 University St., Montr\'eal, Qu\'ebec,  Canada, H3A 2T8.}
\vskip .1in
\centerline{\it ${}^c$ Physics Department, The University of Texas}
\centerline{\it Austin, Texas, USA, 78712.}
\endauthors

\abstract
We study the implications of possible off-peak measurements in the
1995 LEP run, in regard to probing physics beyond the Standard Model.
To do so, we determine the accuracy with which various nonstandard couplings
can
be expected to be measured in the three different scan scenarios
recently discussed by Clarke and Wyatt.  We find that each scan scenario
allows greater sensitivity to a different set of new physics couplings.
Oblique parameters are best measured with the longest scan, while
nonstandard fermion couplings to the $Z^0$ tend to be better
constrained (albeit only marginally) if all of the 1995 LEP measurements are
taken on the $Z^0$ peak.
\endabstract


\vfill\eject
\section{Introduction}

With LEP entering into its final period of measurements on the $Z$ peak,
the question arises as to how to use the remaining time most efficiently.  This
involves a basic tradeoff for all experiments that are run in the vicinity of
the
resonance. On the one hand, longer running on resonance maximizes the
number of produced $Z$'s, and so reduces the statistical errors on all
quantities
which can be measured there. On the other hand, a scan away from the
peak is required to better determine the $Z$ lineshape, and so to more
accurately
determine the $Z$ width.  A decision regarding the relative time to be spent
scanning or running on resonance during the 1995 run is imminent.

\ref\CW{P.E.L. Clarke and T.R. Wyatt, preprint CERN-PPE/94-205 (1994).}

Recently Clarke and Wyatt (CW) \CW\ have quantitatively analyzed
this tradeoff in regard to the expected sensitivity to standard model (SM)
parameters
such as the top mass $m_t$, and the QCD coupling $\alpha_s$. Our intention
in the present note is to similarly analyse the tradeoff that can be expected
in
sensitivity to potentially `new' physics from beyond the SM.  In order to do so
we follow (CW) in comparing the following three scanning scenarios:

\topic{(1) No Scan}
$70 \pbinv$ collected on peak in 1995;
\topic{(2) Intermediate Scan}
$44 \pbinv$ collected on peak and $20
                \pbinv$ collected $\pm 3\GeV$ off peak;
\topic{(3) Long Scan}
$24 \pbinv$ collected on peak and $40
                \pbinv$ collected $\pm 3\GeV$ off peak.
 \endtopic

CW conclude that, although observables such as the forward-backward
asymmetries ---  which improve with better  statistics ---  could be better
determined by running only on peak, a 40 $pb^{-1}$ scan is desirable
due to the improved measurement of the total $Z$ width, or its partial width
into leptons \etc.

\ref\blondel{A. Blondel, preprint CERN-PPE/94-133, August 1994.}
\ref\schaile{D. Schaile, contribution to the proceedings of the 27th
International
Conference on High Energy Physics, Glasgow, July 1994.}

We now turn to the implications for new-physics searches of the above three
scanning scenarios. To do so we perform fits using the following $14$ LEP
observables:
$\AeFB$,  $\AmuFB$, $\AtauFB$, $\Gtot$, $\Sigma$, $\Rel$, $\Rmu$, $\Rtau$,
$\AbFB$, $\AcFB$, $\Rb$, $\Rc$, $\MnsPtau$ and $\AePtau$.
For each of the three scan scenarios, we give, in Table I, the precision with
which we imagine that these observables will have been measured after the
1995 run. In this table we have taken the expected
errors as they are given by CW \CW. For those observables in Table I which are
not directly considered by CW, we have scaled the presently-published LEP
errors
\blondel, \schaile\ by the improvement predicted in ref.~\CW.  As shall become
clear below, we need not specify the central values that would be expected for
these observables after the 1995 run.

\midinsert
$$\vbox{\tabskip=0pt \offinterlineskip
\halign to \hsize{\strut \tabskip=0pt \hfil#\hfil & \hfil#\hfil &\hfil#\hfil
&\hfil#\hfil &\hfil#\hfil &\hfil#\hfil &\hfil#\hfil &\hfil#\hfil &\hfil#\hfil
\cr
\noalign{\hrule}\noalign{\smallskip}\noalign{\hrule}\noalign{\medskip}
$\phantom{XXXXX}$ & \ \ No Scan\ \ & \ \ \ 20 $pb^{-1}$\ \ \ & 40 $pb^{-1}$ &
$\phantom{XXX}$ &$\phantom{XXXX}$ &\ No
Scan \ & \ 20 $pb^{-1}$\ \ & 40 $pb^{-1}$\cr
\noalign{\medskip}\noalign{\hrule}\noalign{\smallskip}
\noalign{\hrule}\noalign{\medskip}
$\AeFB  $ & 0.00153 & 0.00159 & 0.00166 &&$\Rtau   $& 0.033  & 0.033  & 0.035
\cr
$\AmuFB $ & 0.00095 & 0.00098 & 0.00102 &&$\AbFB   $& 0.0016 & 0.0017 & 0.0018
\cr
$\AtauFB$ & 0.00117 & 0.00122 & 0.00127 &&$\AcFB   $& 0.0038 & 0.0041 & 0.0043
\cr
$\Gtot  $ & 0.0029  & 0.0021  & 0.0018  &&$\Rb     $& 0.0013 & 0.0013 & 0.0013
\cr
$\Sigma $ & 0.033   & 0.034   & 0.034   &&$\Rc     $& 0.0064 & 0.0064 & 0.0064
\cr
$\Rel    $ & 0.032   & 0.032   & 0.034  &&$\MnsPtau$& 0.005  & 0.005  & 0.005
\cr
$\Rmu    $& 0.028  & 0.028  & 0.03      &&$\AePtau $& 0.0052 & 0.0055 & 0.0057
\cr
\noalign{\medskip}\noalign{\hrule}\noalign{\smallskip}\noalign{\hrule}
}}$$
\centerline{{\bf Table I}}
\medskip
\centerline{Post-1995 Uncertainties in LEP Observables}
{\eightrm The expected end-of-1995 standard deviations for some LEP observables
considered in this analysis.}
\endinsert

\ref\summary{C.P. Burgess, in the proceedings of the 3rd Workshop on High
Energy Particle Physics, (hep-ph/9411257).}

\ref\STU{M.E. Peskin and T. Takeuchi, \prl{65}{90}{964}; \prd{46}{92}{381}; \bk
W.J. Marciano and J.L. Rosner, \prl{65}{90}{2963}; \bk
D.C. Kennedy and P. Langacker, \prl{65}{90}{2967}; \bk
B. Holdom and J. Terning, \plb{247}{90}{88}.}

\ref\Epses{G. Altarelli and R. Barbieri, \plb{253}{91}{161}; \bk
G. Altarelli, R. Barbieri and S. Jadach, \npb{369}{92}{3}, (erratum)
{\it ibid.} {\bf B376} (1992) 444; \bk
G. Altarelli, R. Barbieri and F. Caravaglios, \npb{405}{93}{3}.}

\ref\VWX{I. Maksymyk, C.P. Burgess and D. London, \prd{50}{94}{529}
(hep-ph/9306267); \bk
C.P Burgess, S. Godfrey, H. K\"onig, D. London and I. Maksymyk,
\plb{326}{94}{276} (hep-ph/9307337); \bk
P. Bamert and C.P. Burgess, {\it Zeit. Phys. C} (to appear),
(hep-ph/9407203).}

\ref\obliquemf{S. Fleming and I. Maksymyk, preprint UTTG-04-95,
NUHEP-TH-95-3}

In order to infer the implications such improved measurements will have for
new-physics searches, we would like to express the implications of such
new physics for these observables in a reasonably model-independent way.
In what follows, we perform two such model-independent
analyses.\foot\bothsummarized{Both
of these methods have been recently summarized and compared in ref.~\summary.}
We first consider the widely-studied
case for which new physics dominantly enters into LEP observables through
the three oblique parameters \STU, \Epses, \VWX, \obliquemf.
In this case the theoretical
predictions for the LEP  observables may be written as a
radiatively corrected standard-model piece
plus a deviation which is linear in the two new physics
parameters $S$ and
$T$. Our goal is to determine the accuracy with which $S$ and $T$ will be
constrained using the three scenarios for the 1995 run. Notice that since the
observables are linear functions of the parameters $S$ and $T$, the
{\it precision} with which $S$ and $T$ will be measured does not depend
on the central values that are assumed to have been found for the observables
after the 1995 run. (The same is not true, of course, for the central values
for $S$ and $T$.)

\ref\bigfit{C.P. Burgess, S. Godfrey, H. K\"onig, D. London and I. Maksymyk,
\prd{49}{94}{6115} (hep-ph/9307337).}

In our second analysis, we drop the assumption that new physics
is dominantly oblique. Rather, we assume only that
the new physics is heavy, so that its effects can be expressed in terms of the
low-dimension
interactions of an appropriate effective lagrangian. This can be thought of as
the lagrangian which would be left after all of the new, heavy particles have
been `integrated out'. The most general expression for such a lagrangian,
subject
to the restriction that it contain only the presently observed particles, is
given
in ref.~\bigfit, which also shows how LEP observables depend on its effective
couplings.  We choose here for simplicity to work with a particularly
interesting
subset of these effective interactions, namely nonstandard couplings between
each flavor of fermion and the $Z$. That is, we focus on effective interactions
of the form:
\label\effinteractions
\eq
\Scl_{\rm eff} =  {ie \over \cw \sw} \; Z_\mu \sum_f \overline{f} \gamma^\mu
(g^f_\ssl \Pl + g^f_\ssr \Pr) f ,
\eeq
where the coupling constants, $g^f_{\sss L,R} = (g^f_{\sss L,R})_{\sss SM}
+ \delta g^f_{\sss L,R}$, are normalized so that  $(g^f_\ssl)_{\sss SM}
= T_{3f} - Q_f \ssq$ and $(g^f_\ssr)_{\sss SM} =  - Q_f
\ssq$.\foot\notation{The
$\ss \delta g^f_{\sss L,R}$ correspond to what was denoted $\ss \delta
\tilde{g}^{ff}_{\sss L,R}$ in ref.~\bigfit.}
We may now determine the precision with which each of the parameters
$\ss \delta g^f_{\sss L,R}$
can be ascertained in each of the three scanning scenarios for the 1995 LEP
run.

\section{Oblique New Physics}

\ref\opal{The OPAL Collaboration, \zpc{61}{94}{19}.}

The precision which we obtain for the two oblique parameters, $S$ and $T$,
when these are fit to the LEP observables using the anticipated experimental
errors from Table I are listed in Table II.  All of the errors listed in this
table represent 2-$\sigma$ allowed ranges.
Results are listed for two kinds of fits.
The columns labelled `Individual Fit' are performed with
only one parameter allowed to vary, the other
parameter being set to zero. By contrast, the `Global Fit' column gives
the result of a full two-parameter maximum-likelihood analysis. The fits were
performed using the correlation matrix given in ref.~\opal\ for the leptonic
observables, and the correlation matrix of ref.~\schaile\ for the heavy-quark
quantities. No correlations were assumed between these two types of
observables.

\midinsert
$$\vbox{\tabskip=0pt \offinterlineskip
\halign to \hsize{\strut \tabskip=0pt \hfil#\hfil & \hfil#\hfil &\hfil#\hfil
&\hfil#\hfil &\hfil#\hfil &\hfil#\hfil &\hfil#\hfil &\hfil#\hfil &\hfil#\hfil
\cr
\noalign{\hrule}\noalign{\smallskip}\noalign{\hrule}\noalign{\medskip}
& & & Individual Fit & & & & Global Fit & \cr
& $\phantom{XXXX}$ & No Scan & 20 $pb^{-1}$ & 40 $pb^{-1}$ &
$\phantom{XXXX}$ & No
Scan  & 20 $pb^{-1}$ & 40 $pb^{-1}$ \cr
\noalign{\medskip}\noalign{\hrule}\noalign{\smallskip}
\noalign{\hrule}\noalign{\medskip}
& $S$ & 0.108 & 0.112 & 0.115 && 0.262 & 0.219 & 0.209 \cr
& $T$ & 0.129 & 0.117 & 0.110 && 0.311 & 0.229 & 0.199 \cr
\noalign{\medskip}\noalign{\hrule}\noalign{\smallskip}\noalign{\hrule}
}}$$
\centerline{{\bf Table II}}
\medskip
\centerline{Oblique Parameter Fit}
{\eightrm The expected sensitivity to electroweak oblique parameters for the
three types of scans considered. All error
ranges indicate 2-$\ss \sigma$ intervals.}
\endinsert

Two points are suggested by the results of Table II.

\item 1
Inspection of the `Global' fit shows that both of the oblique parameters are
better measured in the scenario with a 40 $pb^{-1}$ scan.

\item 2
An opposite conclusion would have been reached for the parameter $S$
if an `Individual' fit had been performed with $T$ constrained to vanish.
This, however, would only be an appropriate fit if the new physics should
be known to be dominated by oblique corrections, and if there are \apriori\
theoretical reasons for $T$ to be much smaller than $S$.

\section{Nonstandard Neutral-Current Fermion Couplings}

We now turn to the case where new physics induces nonstandard neutral-current
couplings for fermions. If we count an independent left- and right-handed
coupling
for each of the eleven light quarks and leptons, then there are potentially 22
couplings to be considered in this case. Happily, not all of these need be
considered
separately, since not all of these enter into the LEP observables in a
linearly-independent
way. In particular, the couplings of the three lightest quarks ($u$, $d$ and
$s$) only
appear through the linear combination
\label\UDdef
\eq
\UD = \sum_{q = u,d,s} \Bigl[ (g^q_\ssl)_{\sss SM} \; \delta g^q_\ssl +
(g^q_\ssr)_{\sss SM} \; \delta g^q_\ssr \Bigr] .
\eeq
We also choose to ignore the six nonstandard neutrino neutral-current couplings
in what follows. This leaves 11 free parameters with which to confront the
data.

When the same two types of fits that were performed earlier for the oblique
parameters are repeated, the results in Table III are obtained. All errors
again
correspond to 2-$\sigma$ ranges.

\midinsert
$$\vbox{\tabskip=0pt \offinterlineskip
\halign to \hsize{\strut \tabskip=0pt \hfil#\hfil & \hfil#\hfil &\hfil#\hfil
&\hfil#\hfil &\hfil#\hfil &\hfil#\hfil &\hfil#\hfil &\hfil#\hfil &\hfil#\hfil
\cr
\noalign{\hrule}\noalign{\smallskip}\noalign{\hrule}\noalign{\medskip}
& & & Individual Fit & & & & Global Fit & \cr
& $\phantom{XXXX}$ & No Scan & 20 $pb^{-1}$ & 40 $pb^{-1}$ &
$\phantom{XXXX}$ & No Scan  & 20 $pb^{-1}$ & 40 $pb^{-1}$ \cr
\noalign{\medskip}\noalign{\hrule}\noalign{\smallskip}
\noalign{\hrule}\noalign{\medskip}
 & $\delta g^e_\ssl$ & 0.000597 & 0.000607 & 0.000643 && 0.00109
& 0.00110 & 0.00113 \cr
 & $\delta g^e_\ssr$ & 0.000603 & 0.000619 & 0.000654 && 0.00113
& 0.00115 & 0.00119 \cr
 & $\delta g^\mu_\ssl$ & 0.000625 & 0.000624 & 0.000667 && 0.00223
& 0.00229 & 0.00238 \cr
 & $\delta g^\mu_\ssr$ & 0.000721 & 0.000721 & 0.000770 && 0.00257
& 0.00264 & 0.00275 \cr
& $\delta g^\tau_\ssl$ & 0.000722 & 0.000722 & 0.000761 && 0.00121
& 0.00119 & 0.00119 \cr
 & $\delta g^\tau_\ssr$ & 0.000811 & 0.000811 & 0.000853 && 0.00133
& 0.00132 & 0.00133 \cr
 & $\delta g^b_\ssl$ & 0.00148 & 0.00138 & 0.00135 &&  0.00658
& 0.00686 & 0.00712 \cr
 & $\delta g^b_\ssr$ & 0.00738 & 0.00702 & 0.00694 && 0.0335
& 0.0352 & 0.0368 \cr
 & $\delta g^c_\ssl$ & 0.00204 & 0.00186 & 0.00182 && 0.0148
& 0.0150 & 0.0152 \cr
 & $\delta g^c_\ssr$ & 0.00446 & 0.00411 & 0.00402 && 0.0189
& 0.0201 & 0.0209 \cr
 & $\delta_{\sss UD}$ & 0.000705 & 0.000645 & 0.000628 && 0.00513
& 0.00510 & 0.00509 \cr
\noalign{\medskip}\noalign{\hrule}\noalign{\smallskip}\noalign{\hrule}
}}$$
\centerline{{\bf Table III}}
\medskip
\centerline{Fermion Coupling Fit}
{\eightrm The expected sensitivity to nonstandard neutral current couplings for
the three types of scans considered. All
error
ranges indicate 2-$\ss \sigma$ intervals.}
\endinsert

Several conclusions emerge from these results.

\item 1
For the `Global' fit, all parameters except $\tauL$, $\tauR$ and $\UD$, were
best
constrained by the no-scan scenario, although the difference between the
no-scan
and the 40 $pb^{-1}$ scan are in many cases not large.

\item 2
In the `Individual' fits,  the heavy-quark couplings become more constrained in
the
40 $pb^{-1}$ scan. However this conclusion is only applicable if there are
reasons
to expect all other couplings to be negligible in a particular model. Since it
is
generically true that most kinds of new physics generate more than one of these
effective couplings at once, we take this as a warning against drawing
meaningful
conclusions from individual fits.

\item 3
The relative improvement or deterioration of the measurement of the
neutral-current
couplings in all three of the scan scenarios appears to be much weaker than was
the
case for the oblique parameters.

\section{Conclusions}

Our goal has been to analyse the implications of the three scanning scenarios
on the precision with which the LEP experiments can be expected to constrain
(or detect!) new physics in their 1995 run. We have done so by parameterizing
the assumed new physics in a relatively model-independent way. We have
considered
two ways of doing so: ($i$) using the oblique parameters, $S$ and $T$, and
($ii$)
using a set of nonstandard neutral-current couplings for all of the known
charged
fermions.

Our conclusions as to the relative efficiency of the the various scan scenarios
are mixed. We have found that when new physics is well described by the oblique
parameters $S$ and $T$, these parameters are best constrained in the scenario
with the longest scan. This is because one can profit from the improved
accuracy
with which the total width and the various leptonic widths are known.  The same
conclusion holds, although more weakly, if the new physics first shows up in
the
neutral-current couplings of the light quarks ($u$, $d$ and $s$) or the tau
lepton.
Otherwise --- for new physics in the other neutral-current couplings --- the
best case is to run continually on resonance. We have also found that whereas
measurements of oblique corrections are reasonably sensitive to which scanning
scenario is used, the same is not true for exotic fermion-$Z$ couplings.

\bigskip

\centerline{\bf Acknowledgements}
\medskip
We would like to thank Dean Karlen for suggesting this line of
inquiry, and to acknowledge the Swiss National Foundation, NSERC
of Canada, FCAR du Qu\'ebec, and the US Department of Energy.

\listrefs

\bye